\documentclass[a4paper,11pt]{article}
\usepackage{amsmath,amsthm,latexsym,amssymb,amsfonts,epsfig}
\usepackage{fontenc,dsfont,layout}
\usepackage{hyperref}
\usepackage[cp1250]{inputenc}
\usepackage{graphicx}
\numberwithin{equation}{section}

\newcommand{\baza}[2]{^o \! e^{#1}_{#2}}
\newcommand{\bazad}[2]{^o \! \omega^{#1}_{#2}}

\DeclareMathAlphabet{\mathpzc}{OT1}{pzc}{m}{it}

\begin{document}

\title{Loop Quantum Cosmology corrections to inflationary models}
\author{Micha{\l} Artymowski\thanks{svetaketu@tlen.pl} $\;\;$
           Zygmunt Lalak\thanks{Zygmunt.Lalak@fuw.edu.pl} $\;\,$  and $\;$
          {\L}ukasz Szulc\thanks{lszulc@fuw.edu.pl}}
\date{\it  Institute of Theoretical Physics, University of Warsaw ul. Ho\.{z}a 69, 00-681 Warszawa, Poland} 
\maketitle

\begin{abstract}
In the recent years the quantization methods of Loop Quantum Gravity have been successfully applied to the homogeneous and isotropic Friedmann-Robertson-Walker space-times. The resulting theory, called Loop Quantum Cosmology (LQC), resolves the Big Bang singularity by replacing it with the Big Bounce. We argue that LQC 
generates  also certain corrections to field theoretical inflationary scenarios. These corrections imply  that in the LQC the effective sonic horizon becomes 
infinite at some point after the bounce and that the scale of the inflationary potential implied by the COBE normalisation increases. 
The evolution of scalar fields immediately after the Bounce becomes modified in an interesting way. We point out that one can use COBE normalisation to establish an upper bound on the quantum of length of LQG.

\end{abstract}


\section{Introduction}
Loop Quantum Cosmology is a mathematically consistent theory of quantum cosmological space-times \cite{Bojo-rev}. Among various possible cosmological models the ones which are best understood are the  Friedmann-Robertson-Walker models \cite{APS,APVS,SKL,Van-open,Szulc-open}. 
In these case it has been  shown that the quantum isotropic and homogeneous gravitational degrees of freedom minimally coupled to the massless scalar field provide non-singular evolution for the flat, closed and open universes. The Bing Bang singularity becomes replaced with 
 the smooth Big Bounce. In this sense the initial (and in the case of the closed universe also the final) cosmological singularity 
is resolved by the quantum gravitational repulsion effects \cite{APS,APVS,Van-open} when the universe enters the Planckian regime. 

The obvious question to ask is the one about possible traces/imprints of the quantum gravitational effects in the well known, 
successful inflationary scenarios\footnote{To learn more about inflationary scenarios and the theory of the scalar perturbations see [\ref{mukhanov}]}. Could such corrections be strong enough to be measured? Is it possible to detect somehow the Big Bounce after the inflation?  Some of these questions were discussed earlier, for instance in \cite{ZL}, for the flat FRW universe on the basis of the effective semi-classical theory developed in \cite{APS}. Although the results like the super-inflation phase without phantom matter are interesting, the potentially observable correction to the primordial perturbation power spectrum was found to be to weak. While, as we will see in
the next sections the predictions coming from the Loop Quantum Cosmology
are indeed rather to week to observe it is possible to find observational
bound to some parameters of the theory. 

This paper is organized as follows. In the section (\ref{sec:loopquantumcosmology}) we briefly discuss General Relativity formulated in Ashtekar variables, and the issue of symmetry reduction  for the case of flat FRW universe. In the section (\ref{LQC-modifi}) the basics of the effective Loop Quantum Cosmology are introduced. In sections (\ref{correct}) and (\ref{initial}) basic differences between classical and LQC effective inflationary cosmologies are described with the help of simple examples. The scalar perturbations are studied in the section (\ref{perturb}). Conclusions are contained in the final section of the paper.

\section{Loop Quantum Cosmology} \label{sec:loopquantumcosmology}
\subsection{Hamiltonian formulation of the classical General Relativity}
In this section we briefly recall the formulation of General Relativity (GR) in Ashtekar variables\footnote{Detailed formulation of GR in Ashtekar variables can be found in \cite{AL-rev,Thiemann,Rovelli}.}. On the globally hyperbolic, Lorentzian, four dimensional manifold with the metric tensor $g$, the topology is set to be $M=\mathbb{R} \times \Sigma$, where $\Sigma$ denotes the spatial slice of the foliation. Let us denote by $q=q_{ab} dx^a dx^b$ the pullback of $g$ to the $\Sigma$, where $x^a$ are some coordinates on the spatial leaf of the foliation. The spatial metric can be decomposed using triads as $q_{ab} = \delta_{ij} e^{i}_{a} e^{j}_{b}$. The canonically conjugated Ashtekar variables are defined then by the vector fields of density weight 1 as
\begin{equation}
E^a_i = \sqrt{|\mathrm{det} q |}  \ e^a_i
\end{equation} 
and $su(2)$ connection 
\begin{equation}
A^i_a = \Gamma^i_a + \gamma K^i_a
\end{equation} 
where the spin connection is $\Gamma^i_a=-\frac{1}{2}\varepsilon^{ijk}e^b_j(2 \partial_{[a}e^k_{b]} + e^c_k e^l_a \partial_c e^l_b )$ and extrinsic curvature $K$ is defined as Lie derivative of $q$ with respect to the normal vector to the spatial slice as $K_a^i = (\pounds_{\vec{n}} q_{ab}) \delta^{ij} e^b_j$. The $\gamma$ is the so-called Barbero-Immirzi parameter. The Poisson Bracket reads as follows
\begin{equation}
\{ A^i_a(x), E^b_j(y) \} = 8\pi G \gamma \delta^b_a \delta^i_j \delta^3(x,y) \ .
\end{equation}
The Hamiltonian for GR is given by the sum of constraints, where the most important one is the scalar constraint defined by
\begin{equation}\label{total-sc}
C_{\mathrm gr}=\frac{1}{16 \pi G} \int_{\Sigma} d^3 x N(x)\Big( \frac{E^a_i
  E^b_j}{\sqrt{|\mathrm{det}E|}} {\varepsilon^{ij}}_k F_{[ab]}^k -
  2(1+\gamma^2) \frac{E^a_i E^b_j}{\sqrt{|\mathrm{det}E|}} K^i_{[a}
  K^j_{b]} \Big) \ ,
\end{equation}
where the curvature 2-form of the $A$ connection is given as usual by $$F^k_{[ab]} = 2\partial_{[a} A^k_{b]} + {\varepsilon^k}_{ij} A^i_{[a} A^j_{b]} \ .$$ In addition, $N(x)$ is called the lapse function. Let us pass on to the symmetry reduction given by the Friedmann-Robertson-Walker (FRW) space-times. The above formulae are simplified drastically when one restrics the considerations to the FRW metric written as
\begin{equation}\label{flatFRW}
q=a^2(t) \delta_{ij} \ \bazad{i}{a} \ \bazad{j}{b} \ dx^a dx^b 
\end{equation}
where the 1-forms $^o \omega^i$ satisfy $\partial_{[a} \ \bazad{i}{b]}=0$. Such a space-time is called the flat FRW universe \footnote{The metric (\ref{flatFRW}) can be written as $q=a(t)^2 [(1-kr^2)^{-1} dr^2 + r^2d\Omega^2]$ for $k=0$. }. The symmetry reduced variables $E$ and $A$ are given by
\begin{equation}
A_a^i = V_0^{-\frac{1}{3}} c \ \bazad{i}{a} \quad E^a_i = p V_0^{-\frac{2}{3}} |{\rm det} \ ^o \omega| \baza{a}{i} \ ,
\end{equation}
where vector field $\baza{a}{i}$ are dual to the 1-forms in (\ref{flatFRW}) as $\bazad{i}{a} \ \baza{a}{j} = \delta_j^i$. The parameter $V_0$ stands for the volume of the fiducial cell, which must be introduced in the case of homogeneous space-times in order to make the integral (\ref{total-sc}) finite. The variables $c$ and $p$ are related with the scale factor $a(t)$ in the (\ref{flatFRW}) as
\begin{equation}
c=\gamma \frac{\dot{a}(t)}{N} V^{\frac{1}{3}}_0 \quad  p=a^2(t) V^{\frac{2}{3}}_0 \ .
\end{equation}
Moreover, the $c$ and $p$ are canonically conjugated
\begin{equation}\label{sym-redu}
\{c,p \}= \frac{8 \pi G \gamma} {3} \ .
\end{equation}

\section{Loop Quantum Cosmology modifications} \label{LQC-modifi}
Let us begin this section with the observation that in general loop quantization methods give us quantum corrections in the gravitational scalar constraint (\ref{total-sc}) to the term $(\sqrt{|\mathrm{det}E|})^{-1} E^a_i E^b_j {\varepsilon^{ij}}_k$ by the so-called Thiemann trick (this is the inverse triad correction). Also, the curvature 2-form $F$ is modified by SU(2) holonomy along suitable loops (see \cite{APS} for details). During the recent years correction of the first type has been studied in the literature (see for instance the papers \cite{MS,MN,CC} in the case of inflationary scenarios) and in the papers \cite{ACS,Szulc}\footnote{In the case of Bianchi Type I it has been shown firstly in \cite{Chiou} that the Big Bounce occurs only with the holonomy corrections, however, only at the level of the effective semi-classical theory. Also in \cite{Bojo-eff} it has been shown that effective semi-classical theory without inverse triad corrections leads to the Big Bounce.} it has been shown that to obtain the most important result, namely the Big Bounce instead of the Big Bang it is sufficient to consider only holonomy modifications in the quantum gravitational scalar constraint. 

The loop quantization of the flat FRW scalar constraint changes the curvature 2-form $F$ in the following way
\begin{equation}
F^k_{ab} =V^{-\frac{2}{3}}_0 {\varepsilon_{ij}}^k c^2 \ \bazad{i}{a} \ \bazad{j}{b} \to V^{-\frac{2}{3}}_0 {\varepsilon_{ij}}^k \frac{\sin^2 (\bar{\mu} c)}{\bar{\mu}^2} \ \bazad{i}{a} \ \bazad{j}{b} \ , 
\end{equation}
where $\bar{\mu}=\sqrt{\Delta_j}/|p|^{1/2}$. The parameter $\Delta_j=l_{j}^{2}$ stands for the eigenvalue of the area operator \cite{AL-area} and can be interpreted as the quantum of the FRW spacetime. Let us focus our attention on the gravitational part of the scalar constraint. The symmetry reduced version of (\ref{total-sc}) is given by
\begin{equation}\label{sc-reduced}
\mathcal{H}_{\rm gr}= -\frac{3}{8 \pi G \gamma^2} \sqrt{|p|} c^2 \ ,
\end{equation}
where we set the lapse function to be $N=1$ \footnote{Because the Gauss and Diffeomorphism constraints are zero, only the scalar constraint has non-zero contribution to the total Hamiltonian. The symmetry reduced scalar constraint then is called the gravitational Hamiltonian of the system in (\ref{sc-reduced}).}. If the inverse triad corrections are ignored, the effective gravitational part of the FRW scalar constraint assumes the following form
\begin{equation}
\mathcal{H}_{\rm eff}= -\frac{3}{8 \pi G \gamma^2} \sqrt{|p|} \frac{\sin^2 (\bar{\mu} c)}{\bar{\mu}^2} \ .
\end{equation}
Let us observe the simple fact: if $\bar{\mu} \to 0$ then $\mathcal{H}_{\rm eff} \to \mathcal{H}_{\rm gr}$ given by (\ref{sc-reduced}).
Let's now consider the modified Hamiltonian, denoted by $\mathcal{H}$, in the form 
\begin{equation}
\mathcal{H}=\mathcal{H}_{\rm eff} + \mathcal{H}_{mat} \ , \label{eq:hamiltonian}
\end{equation}
where $\mathcal{H}_{mat}$ is the matter Hamiltonian\footnote{The similar effective Hamiltonian, with SU(2) holonomy
corrections only,  was studied in \cite{MSS}, where certain  exact
solutions were found for the case of a massless scalar field with
non-zero cosmological constant.}. 

Let the inflationary Universe be dominated by a scalar field with some potential $V(\phi)$. In this case
\begin{equation}
\mathcal{H}_{mat}=\frac{\pi_{\phi}^{2}}{2|p|^{3/2}} + |p|^{3/2}V(\phi)=p^{3/2}\rho\ , \label{eq:hamiltonianmaterii}
\end{equation}
where $\pi_{\phi}=|p|^{3/2}\dot{\phi}$ is the canonical momentum of the scalar field and $\rho=\rho(t)$ is the energy density of the homogeneous scalar field. We derive equations of evolution of $\phi$ and $a$ from the canonical Hamilton equations of motion given by (\ref{sym-redu}) as follows\footnote{In the rest of the paper we set $(8\pi G)^{-1}=1$.}
\begin{equation}
\dot{\xi}=\{\xi,\mathcal{H}\}=\frac{\gamma}{3} \left( \frac{\partial\xi}{\partial c}\frac{\partial \mathcal{H}}{\partial p} - \frac{\partial\xi}{\partial p}\frac{\partial \mathcal{H}}{\partial c} \right) \ , \label{eq:rniehamiltona}
\end{equation}
where $\xi=\xi(c,p)$ is a function on the classical phase space. Moreover we have the following constraint equation 
\begin{equation}
\mathcal{H}=0 \ . \label{eq:H=0}
\end{equation}
This equation comes from the variation of the action with respect to the lapse function $N$. In this paper we assume that $N=1$ in sections [\ref{sec:loopquantumcosmology}-\ref{initial}] and $N=1+\Phi(t)$ in section [\ref{perturb}]. The loop quantization does not change the variables $p$, $\phi$, $\pi_{\phi}$, $\dot{\phi}$ and $\dot{\pi_{\phi}}$. However, it changes the physical interpretation of $c$, which means that $c\neq\gamma\dot{a}$, where we have set $V_0=1$. We already know that $l_{j}$ can be interpreted as the quantum of the length. Moreover, from \cite{AL-area} we know that the spectrum $\Delta_j$ can by simplified as $\Delta_j=4 \pi l_{\rm Pl}^2 \gamma \sqrt{j(j+1)}$, where $j=1/2, 1, 3/2...$. The quantum number $j$ becomes the additional parameter in the theory (let us note that the most popular value of the $j=1/2$, see \cite{APS,APVS,SKL,Van-open,Szulc-open}). The quantum of length then behaves as $l_{j}\propto (\sqrt{j(j+1)})^{1/2}$.
Equation (\ref{eq:H=0}) is now changed to
\begin{equation}
\sin^{2}\big(\frac{cl_{j}}{\sqrt{p}}\big)=\rho\frac{l_{j}^{2}\gamma^{2}}{3} \ . \label{eq:sin^2c}
\end{equation}
To calculate the Hubble parameter let's take $\dot{p}$ from the equation (\ref{eq:rniehamiltona})
\begin{equation}
\dot{p}=-\frac{\gamma}{3}\frac{\partial H}{\partial c}=\frac{2p}{\gamma l_{j}}\sin\big(\frac{l_{j} c}{\sqrt{p}}\big)\cos\big(\frac{l_{j} c}{\sqrt{p}}\big) \ . \label{eq:pzkropka}
\end{equation}
Defining the Hubble parameter as 
\begin{equation}
H=\frac{\dot{a}}{a}=\frac{\dot{p}}{2p} \ , \label{eq:Hubble}
\end{equation}
we obtain from (\ref{eq:sin^2c}) the equation for $H$ [\ref{chinczycy}] 
\begin{equation}
H^{2} = \frac{1}{3}\rho\big(1-\frac{\rho}{\rho_{cr}}\big) \ , \label{eq:H^2}
\end{equation}
where $\rho_{cr}=\frac{3}{l_{j}^{2}\gamma^{2}}$ is a value of the critical energy density. Thus $\rho_{cr}\propto 1/\sqrt{j(j+1)}$ and the maximal value of the critical energy density is $\rho_{cr}\simeq 0.84G^{2}\simeq 5.3\times 10^{2} M_{pl}^{4}\simeq 1.86\times(10^{19}GeV)^{4}$. Equation (\ref{eq:H^2}) corresponds to the first Friedmann equation.  Since $H^{2}\geq 0$, then $\rho\leq\rho_{cr}$, and for $\rho=\rho_{cr}$ we have $H=0$, which means that Universe stops contracting (or expanding) and the Big Bounce occurs. This limit on $\rho$ is not the only one. Since the maximum value of $\sin(x)\cos(x)$ equals $1/2$, from equations (\ref{eq:pzkropka}), (\ref{eq:Hubble}) and (\ref{eq:H^2}) we obtain
\begin{equation}
H^{2}_{max}=\frac{\rho_{cr}}{12} \ , \qquad \rho_{max}=\frac{\rho_{cr}}{2} \ , \label{eq:ograniczenia}
\end{equation}
where $\rho_{max}$ is the energy density, for which $H$ assumes the maximal value. We calculate $\dot{c}$ analogously to (\ref{eq:pzkropka}), which  gives us the LQC second Friedmann equation 
\begin{equation}
\dot{c}=\frac{\gamma}{3}\frac{\partial H}{\partial p}=\sqrt{p}\big[H\frac{c}{\sqrt{p}}-\frac{\gamma}{2}\rho(1+\omega)\big] \ , \label{eq:czkropka}
\end{equation} 
where $\omega$ is the barotropic coefficient. For the pressure $P$ we have then $P=\omega\rho$. One could assume that inflaton is the source of the energy density in the early Universe, which gives $\omega\in[-1,1]$. In principle we could consider $\omega<-1$, but for LQC such a case would be non physical - the energy density would grow with time beyond $\rho_{cr}$. On the other hand, from equations (\ref{eq:pzkropka}) and (\ref{eq:Hubble}) we get
\begin{equation}
c=\frac{\sqrt{p}}{2l_{j}}\arcsin(2H\gamma l_{j}) \ . \label{eq:c}
\end{equation}
After differentiating (\ref{eq:c}) with respect to time  we obtain from (\ref{eq:H^2}) the equation of motion for $c$
\begin{equation}
\dot{c}=Hc+\gamma\sqrt{p}\frac{\frac{\ddot{a}}{a}-H^{2}}{1-2\frac{\rho}{\rho_{cr}}} \ . \label{eq:czkropkabis}
\end{equation}
Comparing (\ref{eq:czkropka}) with (\ref{eq:czkropkabis}) one finally finds
\begin{equation}
\frac{\ddot{a}}{a}=-\frac{1}{2}\rho(1+\omega)(1-2\frac{\rho}{\rho_{cr}})+\frac{1}{3}\rho(1-\frac{\rho}{\rho_{cr}}) \ . \label{eq:friedmann2}
\end{equation}
This is the LQC second Friedmann equation. It is easy to show, that when $l_{j}\rightarrow 0$, which means $\frac{\rho}{\rho_{cr}}\rightarrow 0$, this equation becomes the second Friedmann equation of the standard cosmology.

We already know how $a(t)$ and $H$ evolve, but we are still missing the evolution equation for $\dot{H}$. We know from the equation (\ref{eq:friedmann2}) that \footnote{This equation can be also obtained by differentiating (\ref{eq:H}).} [\ref{chinczycy}]
\begin{equation}
\dot{H}=-\frac{1}{2}\rho(1+\omega)(1-2\frac{\rho}{\rho_{cr}}) \ , \label{eq:Hzkropka}
\end{equation}
which leads to $\dot{H}<0$ for $\rho<\frac{1}{2}\rho_{cr}$, and to $\dot{H}>0$ for $\rho_{cr}>\rho>\frac{1}{2}\rho_{cr}$. This means that at the beginning of the acceleration epoch, when energy density is maximal and equals $\rho_{cr}$ we have the minimal value of $H$, which grows to the maximum at $H^{2}_{max}=\frac{1}{12}\rho_{cr}$ and keeps decreasing later on. The period $\rho<\frac{1}{2}\rho_{cr}$ makes LQC similar to the cosmology without loop corrections.
\par

One should remember that LQC does not change the nature of matter fields, but it does change the coupling between matter and the curvature. Then formulae for matter energy density, pressure, perturbations, equation of state, continuity equation remain unchanged in the presence of the loop correction.

\section{Quantum corrections to the flat FRW model} \label{correct}

Let us analyse the evolution of the scale factor. Since $\rho\propto a^{-3(1+\omega)}$, then in accordance with the equation (\ref{eq:H^2})  we get differential equation for $a(t)$, which after integration gives the result
\begin{equation}
t=\sqrt{\frac{3}{A}}\int\frac{a^{(1+3\omega)/2}}{\sqrt{1-\frac{A}{\rho_{cr}}a^{-3(1+\omega)}}} da \ , \label{eq:t(a)}
\end{equation}
where $A=\rho_{0}a_{0}^{3(1+\omega)}$ is a proportionality coefficient between  $\rho$ and $a^{-3(1+\omega)}$. The final result is
\begin{eqnarray}
a\propto\big[1+(1+\omega)^{2}\frac{3}{4}\rho_{cr}t^{2}\big]^{\frac{1}{3(1+\omega)}} \ , \label{eq:a(t)} \\
H=\frac{t(\omega+1)\rho_{cr}}{2\big[1+t^{2}(\omega+1)^{2}\frac{3}{4}\rho_{cr}\big]} \ , \label{eq:H} \\
\mathcal{N}(t)=\int_{0}^{t}Hdt=\frac{1}{3}\ln\big[1+(\omega+1)^{2}\frac{3}{4}\rho_{cr}t^{2}\big] \ , \label{eq:efoldy} \\
\rho=\rho_{cr}\big[1+t^{2}(\omega+1)^{2}\frac{3}{4}\rho_{cr}\big]^{-1} \ , \label{eq:rho}
\end{eqnarray}
where $\mathcal{N}(t)$ is the number of e-folds. As we can see $a(t)$ has it's minimal, non zero value and it is the even function of time. Thus $a(t)$ behaves in the same way for the expanding and contracting Universe (positive and negative times). What's more, equation (\ref{eq:a(t)}) after the assumption $l_{j}\rightarrow 0$ gives the scale factor of form identical with FRW without the LQC correction. Let us note, that at $t=0$ the size of the Universe is minimal and simultaneously the Hubble parameter vanishes.
\par

\section{Emergence of  the accelerating Universe} \label{initial}

To analyse equation (\ref{eq:friedmann2}) let us define $\omega_{lim}$ as follows  
\begin{equation}
\omega_{lim}=\frac{2(1-\frac{\rho}{\rho_{cr}})}{3(1-2\frac{\rho}{\rho_{cr}})}-1 \ . \label{eq:omegagraniczne}
\end{equation}
One can check from (\ref{eq:friedmann2}) that the Universe accelerates when 
\begin{equation}
\omega>\omega_{lim}\ for\ \rho>\frac{1}{2}\rho_{cr}\atop\omega<\omega_{lim}\ for\ \rho<\frac{1}{2}\rho_{cr} \ .     \label{eq:warunkiinflcji}
\end{equation}
It follows that for very high energies, when $\rho$ is of order of $\rho_{cr}$ the Universe accelerates for almost any equation of state with $\omega>-1$. For example for $\rho>\frac{1}{4}\rho_{cr}$ we have accelerated expansion for $\omega<0$.  

\begin{figure}[h]
\centering
\includegraphics*[height=7cm]{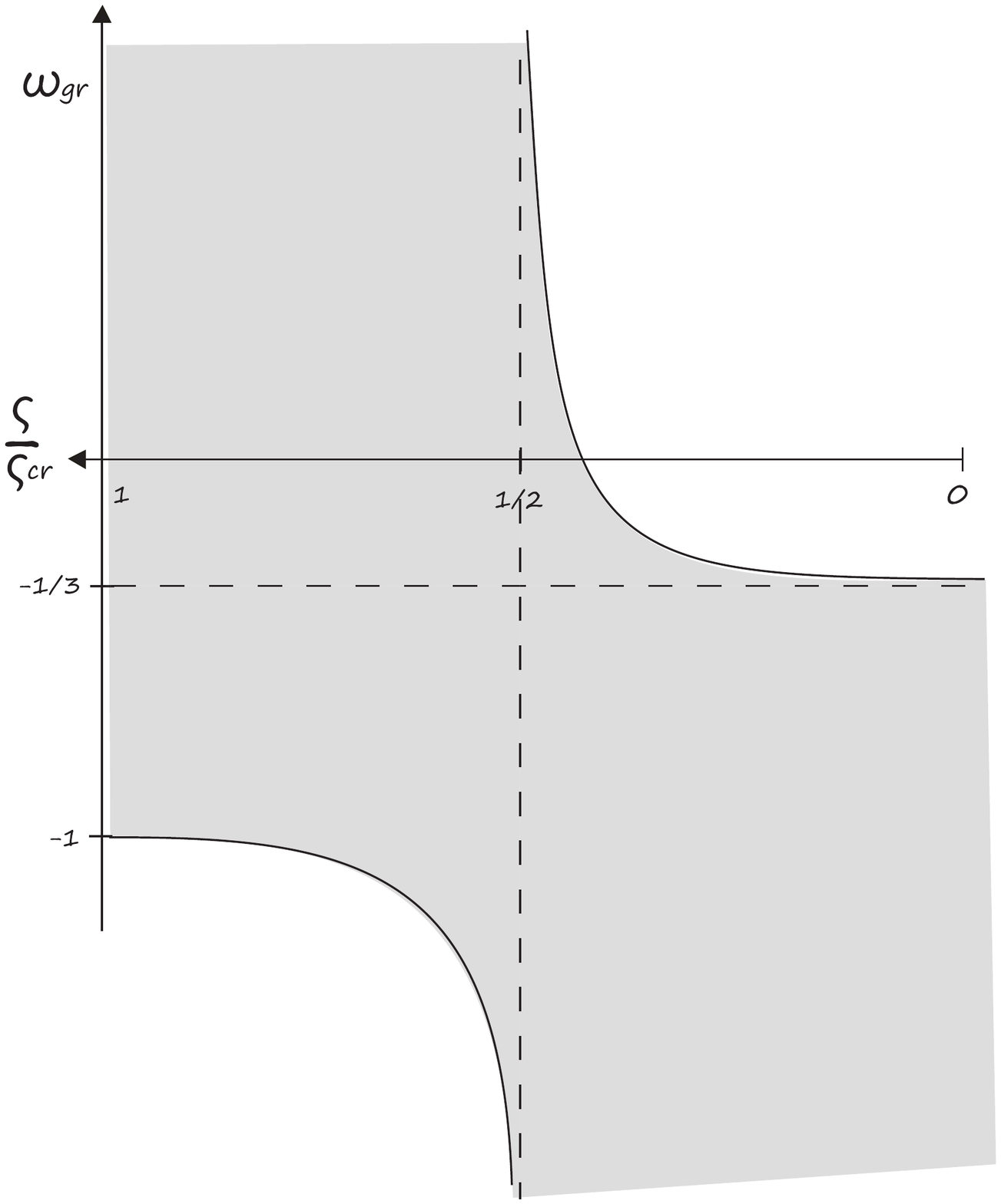}
\hspace{0.5cm}
\includegraphics*[height=7cm]{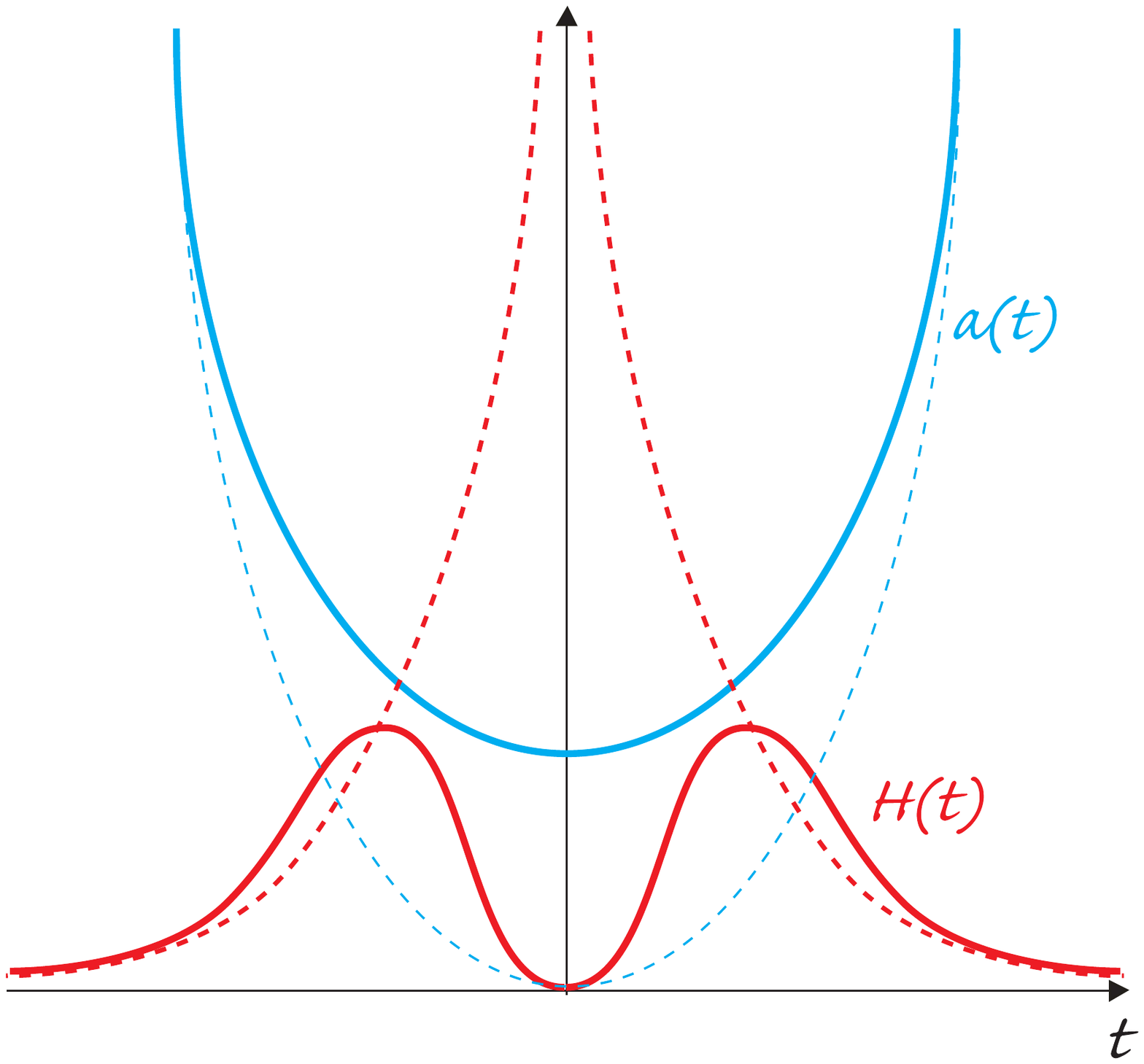}
\caption{\it The left panel  shows $\omega_{lim}$ as a function of $\rho/\rho_{cr}$. Gray area corresponds to accelerated expansion. One can see, that a scalar field produces accelerated expansion  for $\rho>\frac{1}{2}\rho_{cr}$, even without the slow-roll. The right panel shows the evolution of the scale factor and the Hubble parameter in LQC. Dashed lines show their evolution in the standard FRW Universe.}
\label{fig:omegagraniczne}
\end{figure}

\par

From the thermodynamics of the FRW Universe (with or without the loop quantum correction) it is  know that
\begin{equation}
\dot{\rho}+3H(\rho+P)=0 \ . \label{eq:ciaglosci}
\end{equation}
For $\omega=-1$ this gives $\rho=const\Rightarrow a\propto e^{Ht}$, therefore the LQC correction does not significantly alter the time behavior of $a(t)$ for $\omega\simeq -1$. On the other hand  equations of motion for a field and for it's perturbations remain formally unchanged in the presence of the loop correction. Thus in areas of  $\omega\simeq-1$ the power spectrum of field perturbations will be flat for scales much larger than the horizon size. 

\par

In the model with the LQC correction we have two basic differences as compared to the regular FRW. First of all, at the early stages of the evolution of the Universe when $\rho$ is comparable with $\rho_{cr}$, it is natural to obtain an accelerated Universe. Even if initially the would-be inflaton does not lie on the attractor trajectory,
it does produce a period of  accelerated expansion anyway. Secondly the facts that $H$ is bound from above and that $H(t=0)=0$,  significantly limit the initial rate of expansion of the Universe. Unfortunately, even if the Universe accelerates immediately after the Bounce, it is not guaranteed that this initial acceleration will produce enough e-folds to solve problems of classical cosmology. 
As an illustration we shall take the often quoted in the  LQC literature model where close to the big bounce the kinetic energy of a scalar field dominates the energy density, 
 $\rho=\dot{\phi}^{2}/2$. 
\par

Since the LQC correction does not change the equation of motion for the $\phi$, it follows that
\begin{equation}
\ddot{\phi}+3H\dot{\phi}+V'=0 \ , \label{eq:ruchuphi}
\end{equation}
where $V'=\frac{dV}{d\phi}$. For the slow-roll approximation we assume that $\ddot{\phi}\ll 3H\dot{\phi}$ and $\rho\simeq V$, which for the normal FRW and for the potential in the form of $V=\frac{1}{2}m^{2}\phi^{2}$ gives us
\begin{eqnarray}
\phi(t)=\phi_{0}-\sqrt{\frac{2}{3}}mt \ ,  \qquad a\propto\exp\big[\frac{mt}{6}(\phi_{0}\sqrt{6}-m^{2}t)\big] \ . \label{eq:phi(t)s-r} 
\end{eqnarray}
For the FRW Universe with the loop correction one gets from equations (\ref{eq:H^2}), (\ref{eq:ruchuphi}) and slow-roll approximation
\begin{equation}
\sqrt{3V\big(1-\frac{V}{\rho_{cr}}\big)}\frac{d\phi}{V'}=-dt \ . \label{eq:rozniczkowephis-rpetla}
\end{equation}
After taking a specific  form of the potential we can integrate this equation, which for $V=\frac{1}{2}m^{2}\phi^{2}$  (\ref{eq:rozniczkowephis-rpetla}), and 
away from $\rho_{cr}$,  gives
\begin{equation}
t_{0}-t=\sqrt{\frac{3}{4\rho_{cr}}}\Big(\frac{\rho_{cr}}{m^{2}}\arcsin\frac{m\phi}{\sqrt{2\rho_{cr}}}+\frac{\phi}{2}\sqrt{\frac{2\rho_{cr}}{m^{2}}-\phi^{2}}\Big) \ . \label{eq:t(phi)dlaV=phi^2}
\end{equation} 
This reduces to (\ref{eq:phi(t)s-r}) for $m^{2}\phi^{2}\ll\rho_{cr}$. 
\par

As we can see in figure (\ref{fig:omegagraniczne}) a scalar field (or fields), which begins it's evolution with arbitrary initial conditions with $\rho(t=0)=\rho_{cr}$ naturally  produces accelerated expansion. 
Even if the field has big kinetic energy density at $t=0$, it can reach the atractor's trajectory during the $\dot{H}>0$ phase. We would then have successful inflation without initial condition $\dot{\phi}^{2}\ll V(\phi)$. On the other hand, if the inflaton will not reach the atractor's trajectory during the $\dot{H}>0$ phase, we would have a short period of expansion  without inflation. This effect can be seen also in the standard multi-field inflation but it's appearance in the single field inflation is a characteristic feature of LQC. In the standard  FRW cosmology we would expect generation of an acoustic wave during the inflationary pause, which could change the power spectrum of the primordial  energy density perturbations. In LQC we shall not observe this effect because there are no oscillations of the inflaton during the inflationary pause.

How the Loop Quantum Cosmology corrections influence slow-roll parameters? 
The standard FRW Universe accelerates it's expansion when $P<-\frac{1}{3}\rho$, which for the scalar field gives $\frac{1}{2}\dot{\phi}^{2}\ll V$. Equations (\ref{eq:H^2}) and (\ref{eq:ruchuphi}) for slow-roll approximation reduce this inequality to $\epsilon\ll 1$ where 
\begin{equation}
\epsilon=\frac{V'^{2}}{2V^{2}(1-\frac{V}{\rho_{cr}})} \ . \label{eq:epsilon}
\end{equation}
Slow-roll approximation is satisfied when $\ddot{\phi}\ll 3H\dot{\phi}$, which gives us the second slow-roll parameter
\begin{equation}
\eta=\frac{V''}{V(1-\frac{V}{\rho_{cr}})} \ . \label{eq:eta}
\end{equation} 
In the standard  FRW cosmology the inflation appears when both slow-roll parameters are smaller than unity. We will follow this rule for the LQC, because inflation ends when $\rho\ll\rho_{cr}$ and then from the equation (\ref{eq:friedmann2}) we have inflation for $\omega<-\frac{1}{3}$. What's more, for $\rho\ll\rho_{cr}$ the slow-roll parameters are reduced to their original form. The loop correction can be important for $\epsilon$ and $\eta$ only for the COBE normalisation. In the power spectrum of initial energy density perturbations will be proportional to $\dot{\phi}^{2}$  That's why we have decided to define $\epsilon\propto\dot{\phi}^{2}$ and not $\epsilon\propto\dot{H}$. Unfortunately the slow-roll approximation can not be satisfied for the $\rho\simeq\rho_{cr}$, because then $H^{2}\ll\dot{H}$. In such a situation we observe inflation without the slow-roll approximation.
\par

Let us see more detaily how the inflaton joins the attractor's trajectory. If it's initial conditions correspond to the dominance of potential energy, then automatically $\phi$ is on one of the attractor's arms. If initially the kinetic energy dominates, then
\begin{equation}
\ddot{\phi}+3H\dot{\phi}\simeq 0\ , \qquad \rho\simeq\frac{1}{2}\dot{\phi}^{2} \ . \label{eq:dominacjakinetyczna}
\end{equation}
With the initial conditions $\phi(t=0)=0$, $\rho(t=0)=\rho_{cr}$ we get following equations from (\ref{eq:H^2})
\begin{equation}
\dot{\phi}=\sqrt{2\rho_{cr}}\sin\big[{\rm arccot}(\sqrt{3\rho_{cr}}t)\big] \ ,  \qquad
\phi=\sqrt{\frac{2}{3}}\ln\Big[\sqrt{3\rho_{cr}}t+\sqrt{3\rho_{cr}t^{2}+1}\Big] \ . \label{eq:ewolucjaphikin}
\end{equation}
From the above formulae for $\phi$ and $\dot{\phi}$ one can see, that with the flow of time $\phi$ increases rapidly and  $\dot{\phi}$ decreases. This means, that the potential energy gradually begins to dominate, thus the inflaton necessarily reaches the arm of the attractor.

Let us go back for a while to the no-potential energy case, the standard example discussed in most papers on LQG. We have just demonstrated how fast the kinetic energy dominance fails, if only the inflaton has a potential of any kind. 
To satisfy curiosity one can easily  check that the period of acceleration, which begins for $\rho=\rho_{cr}$ will end as foreseen in (\ref{eq:warunkiinflcji}) when $\rho=\frac{2}{5}\rho_{cr}$. Hence it is  extremely short, and  the corresponding number of e-folds is  $0,16$ {}\footnote{To solve problems of the standard cosmology one needs 60-70 e-folds.}.  One necessarily needs a suitable scalar potential to produce 60-70 e-folds.
\par

In what follows we shall show that  the LQC corrections can be described  by standard  FRW Universe with the specific definitions of effective variables. Let us introduce the efetive preassure and energy density
\begin{eqnarray}
\rho_{eff}=\rho(1-\frac{\rho}{\rho_{cr}}) \ , \label{eq:rhoeff} \\
P_{eff}=P(1-2\frac{\rho}{\rho_{cr}})-\frac{\rho^{2}}{\rho_{cr}} \ . \label{eq:Peff}
\end{eqnarray} 
One can verify  that $\rho_{eff}$ and $P_{eff}$ satisfy formally Fiedmann equations with loop corrections omitted. Equations (\ref{eq:rhoeff}-\ref{eq:Peff}) lead directly to the effective barotropic coefficient $\omega_{eff}=P_{eff}/\rho_{eff}$, where 
\begin{equation}
\omega_{eff}=\frac{\omega(1-2\frac{\rho}{\rho_{cr}})-\frac{\rho}{\rho_{cr}}}{1-\frac{\rho}{\rho_{cr}}} \ . \label{eq:omegaeff}
\end{equation}
For large $\rho$ we have $\omega\neq\omega_{eff}$. 
One should note remember that during the slow-roll approximation $\rho\simeq V$ which gives
\begin{equation}
V_{eff}=V(1-\frac{V}{\rho_{cr}}) \ . \label{eq:Veff}
\end{equation}
\par

\section{Scalar Perturbations} \label{perturb}

Let us consider the evolution of perturbations from the Hamiltonian perspective. We shall consider only large scale metric perturbations for which $k\rightarrow 0$. This means that we analyse scales much larger than the comoving sonic horizon, 
which will be defined later. Then the perturbations are  functions of time only. Calculations are much more complicated if one includes  also explicit  $\vec{x}$ dependence. By continuity one can argue that the limit of $k\rightarrow 0$ of the full calculation 
shall give us the same results. 

Let us start with the derivation of the equations of motion for perturbations without the loop correction.  According to the definition of the canonical variable $p$ for the line element 
\begin{equation}
ds^{2}=-N^{2}dt^{2}+a^{2}(1-2\Psi)(d\vec{x})^{2}\ ,  \label{eq:metrykaperturbowana} 
\end{equation}
where $N=1+\Phi(t)$, one  obtains 
\begin{equation}
p=a^{2}\big(1- 2\Psi(t)\big) \ . \label{eq:pperturbowane}
\end{equation}
The $\Phi$ and $\Psi$ are metric scalar perturbations in the Newtonian gauge. We can now express equation (\ref{eq:metrykaperturbowana}) in the form 
\begin{equation}
ds^{2}=-N^{2}dt^{2}+p(d\vec{x})^{2}. \label{eq:metrykaperturbowanazp}
\end{equation}
This results, via Hamilton equations,  in Friedmann equations for metric perturbations in the form which agrees with the reference  [\ref{mukhanov}]
\begin{eqnarray}
3H\dot{\Psi}+3H^{2}\Phi=-\frac{1}{2}\delta\rho \ , \label{eq:friedmann1pert} \\
\ddot{\Psi}+3H\dot{\Psi}+H\dot{\Phi}+(2\dot{H}+3H^{2})\Phi=\frac{1}{2}\delta P \ . \label{eq:friedmann2pert}
\end{eqnarray}

For the scalar field  the energy-momentum tensor is diagonal, which is not true for the Einstein curvature tensor. The scalar field  $\phi$ does not produce anisotropic stress, which leads  to  $\Phi=\Psi$. Since $\delta P=c_{s}^{2}\delta\rho+\tau\delta S$, where $\delta S$ is a perturbation of the entropy density, we obtain the adiabatic equation of motion for $\Phi$ [\ref{mukhanov}]
\begin{equation}
\ddot{\Phi}+H(4+3c_{s}^{2})\dot{\Phi}+(2\dot{H}+3H^{2}(1+c_{s}^{2}))\Phi=0 \ , \label{eq:ruchuPhi}
\end{equation}
where  
\begin{equation}
c_{s}^{2}=\Big(\frac{\partial P}{\partial\rho}\Big)_{_{S}}=\frac{\delta P}{\delta\rho}  \label{eq:c_sFRW}
\end{equation}
is the square of the speed of sound calculated with  constant entropy. The solution of equation (\ref{eq:ruchuPhi}) (we consider  scales much larger than the sonic horizon,   $kc_{s}^{2}\ll aH$) with constant $\omega$ reaches  very fast a time independent value determined by the value of $\omega$. 

Taking the loop correction into account  and following the same procedure as  for the standard FRW one derives  equations analogous to  (\ref{eq:friedmann1pert}-\ref{eq:ruchuPhi}). The equations of motion of scalar metric perturbations in LQC 
in the limit  $k\rightarrow 0$ read
\begin{eqnarray}
3H\dot{\Psi}+3H^{2}\Phi=-\frac{1}{2}\delta\rho_{eff}=-\frac{1}{2}\delta\rho(1-2\frac{\rho}{\rho_{cr}}) \ , \label{eq:friedmann1pertpetla} \\
\ddot{\Psi}+3H\dot{\Psi}+H\dot{\Phi}+(2\dot{H}+3H^{2})\Phi=\frac{1}{2}\delta P_{eff}=\frac{1}{2}\delta P(1-2\frac{\rho}{\rho_{cr}})-2\frac{\rho+P}{\rho_{cr}}\delta\rho \ , \label{eq:friedmann2pertpetla} \\
\ddot{\Phi}+H(4+3c_{s_{eff}}^{2})\dot{\Phi}+(2\dot{H}+3H^{2}(1+c_{s_{eff}}^{2}))\Phi=0 \ , \label{eq:ruchuPhieff}
\end{eqnarray}
where $c_{s_{eff}}$ is the effective speed of sound defined by
\begin{equation}
c_{s_{eff}}^{2}=\frac{\delta P_{eff}}{\delta\rho_{eff}}=\frac{\delta P}{\delta\rho}-2\frac{\rho+P}{\rho_{cr}-2\rho}=c_{s}^{2}-2\frac{\rho+P}{\rho_{cr}-2\rho} \ . \label{eq:c_s}
\end{equation}
In the calculations given above we have assumed a constant value of the entropy density. 

Let us note that the $c_{s_{eff}}^{2}$ becomes infinite for $\rho=\rho_{cr}/2$. 
This is a novel  effect, a special feature of the loop corrected FRW. Such behaviour is absent in other models with a bounce, like the ekpyrotic Universe [\ref{Ovurt},\ref{Steinhardt}] inspired by string theory. However, one should remember, that $c_{s_{eff}}^{2}$ is an effective variable, not necessarily a physical one. What actually happens is a rapid growth of the effective comoving sonic horizon defined by $\frac{c_{s_{eff}}^{2}}{aH}$. 
To be more specific note, that  for  $\rho \sim \rho_{cr}/2$ and  perturbations outside the effective sonic horizon one finds
\begin{equation}
2\dot{H}+3H^{2}(1+c_{s_{eff}})=-\frac{\rho}{\rho_{cr}}\frac{\rho+P}{1-2\rho/\rho_{cr}},
\end{equation}
which is negligible only if  $P=-\rho$, or $\rho\neq\rho_{cr}/2$ (for $\omega=-1+3\epsilon$ or when $\rho_{cr}\ll M_{pl}^{4}$) or if $\rho\ll\rho_{cr}$. One can see, that indeed, perturbations are not frozen outside the horizon when $\rho\sim\rho_{cr}/2$. 
This implies, that  amplitudes of modes with all relevant wavelengths were decreasing during that period. However, 
the net effects of this suppresion is model dependent. If the model stays sufficiently long near $\rho \approx \rho_{cr}/2$, as expected in simple models with very flat potentials, information about the Universe described by negative time would became practically erased over a wide range of scales. 
Thus the FRW Universe with the loop correction might bear some similarity to the standard big bang theory, because all correlations borne before the Bounce could become significantly suppresed. 

Once again we stress, that we work in the picture where  the loop correction does not change the Einstein curvature tensor, or the curvature itself, but it does modify $T^{\mu\nu}$ by changing the pressure and the energy density to their effective counterparts. One could ask, if the effective $T^{\mu\nu}$ is diagonal and if we have a physical perfect fluid of any kind giving rise to such a form of the energy-momentum tensor. The answer seems to be negative since $\frac{\partial P_{eff}}{\partial\dot{\phi}}\dot{\phi}-P_{eff}\neq\rho_{eff}$.  The relative magnitudes  of tensor and scalar perturbations are proportional to the effective speed of sound calculated in the moment of the sonic horizon crossing. If we only could measure primordial gravitational waves, we would know their power spectrum, which would allow us to calculate $c_{s_{eff}}$. 

On the other hand, we could modify curvature to get correct LQC Einstein equations for $T^{\mu\nu}$ made out of a regular  perfect fluid. Thus instead of $\rho_{eff}$ and $H$ we would have $\rho$ and $H_{eff}$. Unfortunately, such an approach  leads to very complicated calculations, being also somewhat  unintuitive. 
\vspace*{0.3cm}

Let's investigate in detail the role of the loop correction in the primordial energy density power spectrum. As shown in section (\ref{sec:loopquantumcosmology}) equations of motion in LQC for both the inflaton and it's perturbations  formally coincide with those obtained in the case of  the stndard FRW Universe. Hence for the FRW with the loop correction we obtain 
\begin{equation}
\mathcal{P}_{\mathcal{R}_{loop}}=\frac{V_{eff}}{24\pi^{2}\epsilon}=(1-\frac{V}{\rho_{cr}})^{2}\mathcal{P}_{\mathcal{R}} \ , \label{eq:powerspectra}
\end{equation} 
where $\mathcal{P}_{\mathcal{R}}$ is the standard expression for the FRW curvature perturbation. We compute  $V$ and 
$\epsilon$ at the instant $kc_{s}=aH$, which is the moment of the sonic horizon crossing. From the COBE normalisation of the CMB anisotropy one obtains 
\begin{equation}
V^{1/2}(1-\frac{V}{\rho_{cr}})/\epsilon^{1/2}=7.7 \, \times 10^{-4}
\end{equation}
60 e-folds before the end of inflation. The loop correction rises the energy density corresponding to the maximal value of $\epsilon=1$. This effect can be important only for relatively small values of $\rho_{cr}$. We could even imagine effective low scale inflation with a very  small $\rho_{cr}$ (which means very large  $j$) and relatively large $V$. The effective energy density is always smaller than the physical one. 
If $j=1/2$ we cannot  see any effect caused by the loop correction in the power spectrum, because $10^{16}GeV\ll M_{Pl}$. 
However, we can look at this issue differently and ask the question about the upper limit on $j$ which results from the requirement that $\rho_{cr}$ should not modify the power spectrum. If (in the crude approximation) $V^{1/4}\sim 10^{16}GeV$ 60 e-folds before the end of inflation, during the horizon exit, then $\rho_{cr}^{1/4}>10^{16}GeV$ implies $j<10^{12}$. The value of $j=10^{12}$ means, that $l_{j}\sim 10^{-25}\, {\rm cm}$, which is still  an undetectable quantum of length. In terrestrial experiments one can probe 
the Newton law down to distances of $10^{-3}$ cm, hence even a very large value of $f$ doesn't contradict existing data.
More precise calculations gives us the limits for specific models. For example for $V=m^{2}\phi^{2}/2,\ m=10^{-5}$ we would have $\rho_{cr}\sim(1.3\times 10^{17}GeV)^{4}$ which gives $j\simeq 3\times 10^{7}$.

Since slow-roll parameters are changed by the LQC, we should remember that the scalar spectral index, in slow-roll regime $n_{s}\simeq 1+2\eta-6\epsilon$, is also modified by the loop quantum correction. We know from the observations, that $n_{s}\simeq 0,96\;$ \cite{Komatsu:2008hk}. The LQC correction modifies the scalar spectral index in following way 
\begin{equation}
n_{s}-1 \rightarrow\frac{n_{s}-1}{1-V/\rho_{cr}} \ .
\end{equation}
For $V\simeq\rho_{cr}/2$ we have $n_{s}\rightarrow 2n_{s}-1$. Hence in a model with $n_s <1$ in the standard FRW, one obtains a smaller value of the index in its  LQC version.  

\vspace*{0.3cm}

We wish to point out again, that one needs such inflationary scenarios, which give more than 60 e-folds after the the sonic horizon becomes infinite. We count only such perturbations, which freeze  after the horizon crossing and come back inside the horizon after inflation. One can see, that perturbations which are crossing the sonic horizon when $\rho>\rho_{cr}/2$ are coming back when 
$\rho\simeq\rho_{cr}/2$. In fact inflation, in the  strict meaning of this word, starts after $\rho=\rho_{cr}/2$. This conclusion is somewhat  unintuitive, as one could imagine the Universe dominated by the scalar field potential energy right from the outset. 
Thus for $\omega\simeq const=-1+3\epsilon$ \footnote{This assumption is fulfilled in the slow-roll approximation, when $\epsilon,\ \eta\ll 1$ and $\dot{\epsilon}=2H\epsilon(2\epsilon-\eta)\ll H \epsilon$.} we have $N(t_{1/2})=\frac{\ln[2]}{\epsilon}$ which can be a very large  number for a very small $\epsilon$. Thus we could produce 70 e-folds during $\dot{H} > 0$ era, but these e-folds do not count from the point of view of the initial energy density perturbations power spectra. To reach this conclusion one doesn't need any specific value of $j$.  

\section{Conclusions}

We have analysed loop quantum cosmology corrections to the standard FRW inflationary cosmology. There are significant differences between LQC and standard FRW scenarios in the energy range close to the critical energy density which appears in LQG. The corrections imply  that in the LQC the effective sonic horizon becomes 
infinite at some point after the bounce which in certain models may suppress the memory of the correlations borne before the Big Bounce (more precisely - before the point of time corresponding to  $\rho=\rho_{cr}/2$). The power spectrum of scalar fluctuations becomes modified in such a way that the   
the scale of the inflationary potential implied by the COBE normalisation increases with respect to the standard FRW case. 
In addition, the value of the scalar spectral index typically decrases with respect to the value predicted by a given model in the FRW 
Universe. We point out that one can use COBE normalisation to establish an upper bound on the quantum of length of LQG, which turns out to be  $l_{j}\sim 10^{6}l_{Pl}$.
Also, the evolution of scalar fields immediately after the Bounce becomes modified in an interesting way. 

The necesary condition to make all these effects observable is a sufficiently small value of the critical energy density i.e. sufficiently 
large LQG quantum of length. 
The most popular theory of the FRW with loop corrections assumes $j=1/2$ which makes $\rho_{cr}$ nearly Planckian. In such a case  the LQC effect in power spectra will not be observed. However, since rising the value of $j$ even by several decades doesn't lead to a contradiction with experiments, it is justified and amusing to pursue possible low-energy/late time  consequences of LQC.

\section{Acknowledgments}
Authors thank Krzysztof Turzy\'nski for numerous helpful comments and suggestions. \\
{\L}. Szulc was partially supported by the Polish Ministry of Science and Higher Education, grant N N202 007734 and Foundation for Polish Science, "Master" grant. This work was partially supported by the EC 6th Framework
Programme MRTN-CT-2006-035863, by the  grant MNiSW  N202 176 31/3844
and  by TOK Project  MTKD-CT-2005-029466.
ZL thanks APC Paris for hospitality.

\end{document}